\documentclass[aps,prl,doublespace,groupedaddress,preprint,showpacs,showkeys]{revtex4}

\usepackage{graphicx}
\begin{document}


\title{Transport Reversal in a Thermal Ratchet}



\author{Baoquan  Ai$^{1,2}$}\author{Liqiu Wang$^{1}$}\author{Lianggang
Liu$^{3}$}

\affiliation{$^{1}$ Department of Mechanical Engineering, The University of Hong Kong, Pokfulam Road, Hong Kong\\
$^{2}$School of Physics and Telecommunication Engineering, South
China Normal University, GuangZhou, China\\
 $^{3}$ Department of Physics, ZhongShan
University, GuangZhou, China}



\begin{abstract}
\baselineskip 0.4 in
\indent Transport of a Brownian particle
moving in a periodic potential is investigated in the presence of
symmetric unbiased external force. The viscous medium is
alternately in contact with the two heat reservoirs. We present
the analytical expression of the net current at quasi-steady state
limit. It is found that the competition of the asymmetric
parameter of potential with the temperature difference leads to
the phenomena like current reversal.  The competition between the
two driving factors is a necessary but not a sufficient condition
for current reversals.
\end{abstract}

\pacs{05. 40. -a, 05. 70. -a, 87. 10. +e}

\maketitle

\baselineskip 0.4 in
\section {1. Introduction}
\indent Transport phenomena play a crucial role in many processes
from physical, biological to social systems. There has been an
increasing interest in transport properties of nonlinear systems
which can extract usable work from unbiased nonequilibrium
fluctuations \cite{1,2,3,4}. This comes from the desire of
understanding molecular motors \cite{5}, nanoscale friction
\cite{6}, surface smoothening \cite{7}, coupled Josephson
junctions \cite{8}, optical ratchets and directed motion of laser
cooled atoms \cite{9}, and mass separation and trapping schemes at
microscale \cite{10}.

\indent The focus of research has been on the noise-induced
unidirectional motion over the last decade. A ratchet system is
generally defined as a system that is able to transport particles
in a periodic structure with  nonzero macroscopic velocity in the
absence of macroscopic force on average. In these systems,
directed Brownian motion of particles is generated by
nonequilibrium noise in the absence of any net macroscopic forces
and potential gradients. Typical examples are rocking ratchets
\cite{4,11}, flashing ratchets \cite{12}, diffusion ratchets
\cite{13}, correlation ratchets \cite{4,14} and white-shot-noise
ratchets \cite{2}. In all these studies, the potential is taken to
be asymmetric in space. It has also been shown that a
unidirectional current can also appear for spatially symmetric
potentials if there exits an external random force either
asymmetric or spatially-dependent. If spatially periodic
structures are exposed to additive Poissonian white shot noise of
zero average, a macroscopic current occurs even in the absence of
spatial asymmetry \cite{2}.

\indent The current reversal is very important in new particle
separation devices such as electrophoretic separation of
micro-particles \cite{15}.  It is also of interest in biology
\cite{a}. Motions of macromolecules are probably responsible for
the vesicle transport inside eukaryotic cells. A typical example
is the motion of proteins along a microtubule, modelled usually by
a ratchet \cite{29}. It is well known that the two typical
proteins, kinesins and dyneins, move along tubulin filaments in
opposite directions. This can be explained by the current
reversal.

\indent The current reversal in ratchet systems can be engendered
by varying system parameters
\cite{16,17,18,19,20,21,22,23,24,25,26,27,28}. The current can be
reversed, for example, by a noise of Gaussian force with non-white
power spectrum in present of stationary periodic potential
\cite{19}. The current reversal can also be obtained in two-state
ratchets if the long arm is kinked \cite{20}. Bier and Astumian
\cite{21} have also found the current reversal in a fluctuating
three-state ratchet.  In the presence of a kangaroo process as the
driving force, the current reversal can be triggered by varying
the noise flatness, the ratio of the fourth moment to the square
of the second moment\cite{22}. The current reversal can be induced
by both an additive Gaussian white and an additive
Ornstein-Uhlenbeck noise in a correlation ratchet \cite{23}. The
current reversal also appear in forced inhomogeneous ratchets
\cite{17,18}.

\indent The previous works on the current reversal are limited to
case of one heat reservoir. The present study extends the study of
current reversal to the case of two heat reservoirs. When a
positive driving factor competes with a negative one, the current
may reverse its direction. The competition between the competitive
driving factors is necessary but not sufficient for the current
reversal. Our emphasis is on finding conditions of generating
current reversal. This is achieved by using a quasi-steady state
limit to solve the Fokker-Planck equation.

\section {2. Net current of the thermal ratchet}
\indent Consider a Brownian particle moving in a sawtooth
potential with an unbiased external force where the medium is
alternately in contact with the two heat reservoirs. This model
was first proposed to describe molecular motor in biological
system \cite{29}. The particle motion satisfies the dimensionless
Langevin equation of motion \cite{30,31}
\begin{equation}\label{1}
m\frac{d^{2}x}{d t^{2}}=-\beta \frac{d x}{d t}-\frac{d U(x)}{d
x}+F(t)+\sqrt{2k_{B}T(x)\beta}\xi(t),
\end{equation}
where $x$ stands for the position of Brownian particle, $m$ the
mass of the particle, $\beta$ the viscous friction drag
coefficient, $k_{B}$ Boltzmann constant, $T(x)$ absolute
temperature. $\xi(t)$ is a randomly-fluctuating Gaussian white
noise of zero mean and the autocorrelation function
$<\xi(t)\xi(s)>=\delta(t-s)$. Here $<...>$ denotes an ensemble
average over the distribution of the fluctuating forces $\xi(t)$.
$F(t)$ is an external periodic force (Fig. 1b), satisfy
\begin{equation}\label{}
F(t+\tau)=F(t), \int^{\tau}_{0}F(t)dt=0.
\end{equation}

The geometry of symmetric potential $U(x)=U(x+L)$ is displayed in
Fig. 1a and $U(x)$ within the interval $0\leq x \leq L$ is
described by
\begin{equation}\label{}
U(x)=\left\{
\begin{array}{ll}
    \frac{U}{L_{1}}x, & \hbox{$0\leq x< L_{1}$;} \\
    \frac{U}{L_{2}}(L-x), & \hbox{$L_{1}\leq x\leq
         L$,} \\
\end{array}
\right.
\end{equation}
where $L=L_{1}+L_{2}$ is the period of the potential. The
temperature $T(x)$ has the same period as the potential $U(x)$.
Therefore, $T(x)=T(x+L)$,
\begin{equation}\label{}
T(x)=\left\{
\begin{array}{ll}
    T+\delta, & \hbox{$0\leq x< L_{1}$;} \\
    T, & \hbox{$L_{1}\leq x\leq
         L$.} \\
\end{array}
\right.
\end{equation}

\begin{figure}[htbp]
  \begin{center}\includegraphics[width=8cm,height=8cm]{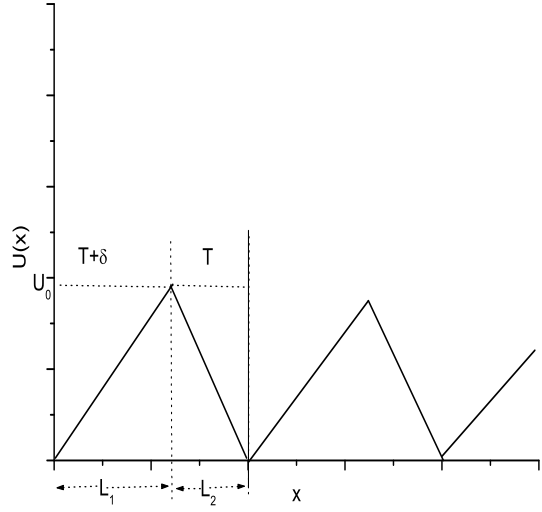}
  \begin{center}(a)\end{center}
  \includegraphics[width=8cm,height=8cm]{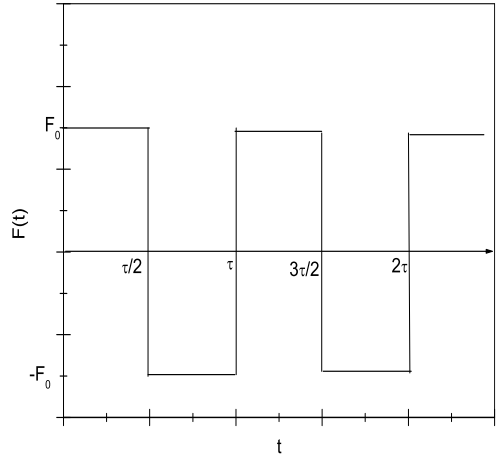}\begin{center}(b)\end{center}
  \caption{\baselineskip 0.4 in Potential and Driving
  force: (a)potential $U(x)=U(x+L)$;
$U(x)$ is a piecewise linear and periodic potential; the period of
the potential is $L=L_{1}+L_{2}$; $\Delta=\L_{1}-\L_{2}$; the
temperature profiles is also shown. (b) Driving force F(t) which
preserved the zero mean $<F(t)>=0$; $F(t+\tau)=F(t)$; $F_{0}$ is
amplitude of $F(t)$.}\label{1}
\end{center}
\end{figure}
\indent Because the motion of the ratchet is highly overdamped in
general \cite{30},
 the inertia term can be neglected. Hence, Eq. (1) reduces to, when $\beta=1$ and
 $k_{B}=1$,
\begin{equation}\label{1}
\frac{d x}{d t}=-\frac{d U(x)}{d x}+F(t)+\sqrt{2T(x)}\xi(t).
\end{equation}

The probability density satisfies the associated Fokker-Planck
equation \cite{30,31},
\begin{equation}\label{2}
\frac{\partial P(x,t)}{\partial t}=\frac{\partial}{\partial
    x}[(U^{'}(x)-F(t))P(x,t)+\frac{\partial}{\partial x}(T(x)P(x,t))]=-\frac{\partial j(x,t)}{\partial
    x},
\end{equation}

\begin{equation}\label{}
    j(x,t)=-[U^{'}(x)-F(t)]P(x,t)-\frac{d}{dx}[T(x)P(x,t)],
\end{equation}
here $j$ is the probability current density. The prime stands for
the derivative with respect to the space variable $x$. $P(x,t)$ is
the probability density for the particle at position $x$ and at
time $t$. It satisfies the normalization condition and the
periodicity condition,

\begin{equation}\label{}
    P(x,t)=P(x+L,t),
\end{equation}

\begin{equation}\label{}
    \int_{0}^{L}P(x,t)dx=1.
\end{equation}

\indent If $F(t)$ changes very slowly with respect to $t$, namely,
its period is longer than any other time scale of the system,
 there exists a quasi-steady state. In this case, by following the  method in  \cite{30,31}, we can obtain the
current $j(F(t))$ from Eq. (7)-Eq. (9),

\begin{equation}\label{}
    j(F(t))=\frac{-Q}{G_{1}G_{2}+HQ},
\end{equation}
where $Q$, $G_{1}$, $G_{2}$ and $H$ are
\begin{eqnarray}
\nonumber
  Q &=& e^{a-b}-1, \\
  G_{1}&=& \frac{L+\Delta}{2a(T+\delta)}(1-e^{-a})+\frac{L-\Delta}{2bT}e^{-a}(e^{b}-1), \\\nonumber
  G_{2}&=&
  \frac{L+\Delta}{2a}(e^{a}-1)+\frac{L-\Delta}{2b}e^{a}(1-e^{-b}),\\\nonumber
  H&=&A+B+C,
\end{eqnarray}

\begin{eqnarray}
\nonumber
  A&=& \frac{1}{T+\delta}(\frac{L+\Delta}{2a})^{2}(a+e^{-a}-1),
  \\\nonumber
  B&=&\frac{L^{2}-\Delta^{2}}{4abT}(1-e^{-a})(e^{b}-1), \\
  C&=&\frac{1}{T}(\frac{L-\Delta}{2b})^{2}(e^{b}-1-b),
\end{eqnarray}
\begin{eqnarray}
\nonumber
   a&=&\frac{2U_{0}-F(t)(L+\Delta)}{2(T+\delta)},\\
   b&=&\frac{2U_{0}+F(t)(L-\Delta)}{2T}.
\end{eqnarray}
The average current is
\begin{equation}\label{4}
J=\frac{1}{\tau}\int_{0}^{\tau}j(F(t))dt,
\end{equation}
where $\tau$ is the period of the driving force $F(t)$. $\tau$ is
assumed to be longer than any other time scale of the system at
quasi-steady state. For the external force $F(t)$ shown in Fig.
1b,
\begin{equation}\label{}
    J=\frac{1}{2}[j(F_{0})+j(-F_{0})].
\end{equation}

\indent When both the potential and the temperature are symmetric
($\delta=0, F_{0}=0$), the current $J$ reduces to:
\begin{equation}\label{}
    J=\frac{1}{2(2T+\delta)}(\frac{U_{0}}{2L})^{2}(\frac{1}{e^{\frac{U_{0}}{T+\delta}}-1}-\frac{1}{e^{\frac{U_{0}}{T}}-1}).
\end{equation}
\indent Therefore, the net current is not zero even when both the
potential and the temperature are spatially symmetric. The
direction of the current is determined by the sign of $\delta$.
The particle tends to move from high temperature region to low
temperature region. In fact, this agrees with the diffuse law.

\section {3. Results and discussion}

\begin{figure}[htbp]
  \begin{center}\includegraphics[width=10cm,height=8cm]{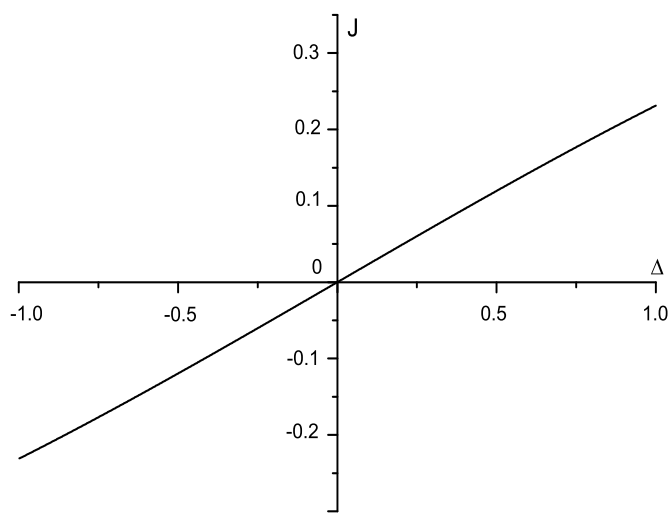}
  \caption{\baselineskip 0.4 in Current $J$ versus asymmetric parameter $ \Delta$ of the potential at $U_{0}=5$, $F_{0}=3.0$, $L=1.0$, $T=1.0$ and $\delta=0$.}\label{1}
\end{center}
\end{figure}
\indent

Figure 2 shows the current $J$ as a function of the asymmetric
parameter $\Delta$ of the potential at $\delta=0$. The current is
negative for $\Delta<0$, zero at $\Delta=0$ and positive for
$\Delta>0$. Therefore, we can have the current reversal by
changing the sign of $\Delta$, the asymmetry of the potential.

\begin{figure}[htbp]
  \begin{center}\includegraphics[width=10cm,height=8cm]{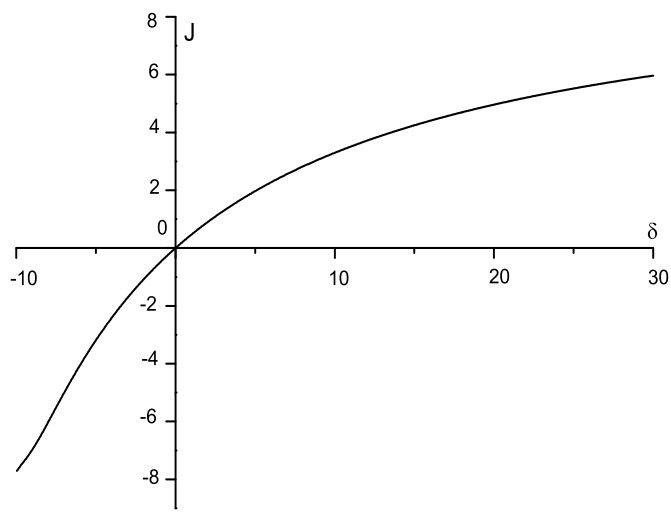}
  \caption{\baselineskip 0.4 in Current $J$ versus temperature difference $\delta$ at $U_{0}=5$, $F_{0}=3.0$, $L=1.0$, $T=10$ and $\Delta=0$. }\label{1}
\end{center}
\end{figure}
\indent  Figure 3 shows the current $J$ versus temperature
difference $\delta$ in a symmetric potential $\Delta=0$. The
temperature difference $\delta$ controls not only the magnitude
but also the direction of the current. When $\delta=0$ and
$\Delta=0$, there is no current. For the asymmetric potentials,
varying temperature difference is another way of inducing a net
current.

\begin{figure}[htbp]
  \begin{center}\includegraphics[width=10cm,height=8cm]{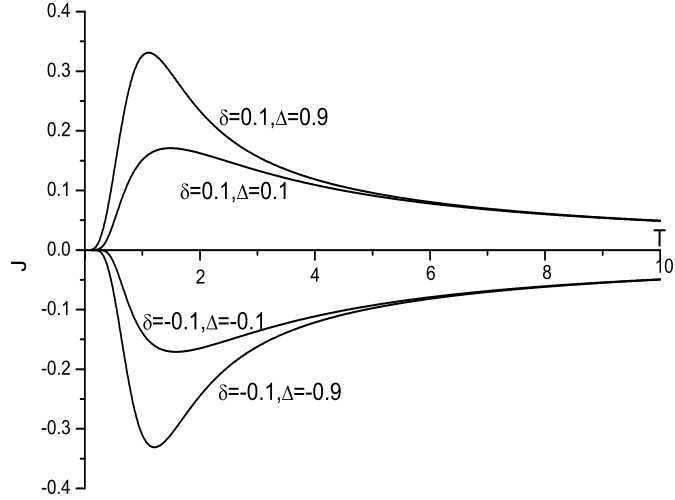}
  \caption{\baselineskip 0.4 in Current $J$ versus temperature $T$ for different asymmetric parameters $\delta$ and$\Delta $ at
  $U_{0}=5$, $F_{0}=3.0$ and $L=1.0$.}\label{1}
\end{center}
\end{figure}

\indent The current $J$ as the function of $T$ is shown in Fig. 4
for different combinations of $\Delta$ and $\delta$. The curve is
observed to be bell-shaped, which shows the feature of resonance.
When $T\rightarrow 0$, $J$ tends to zero for all values of
$\delta$ and $\Delta$.  Therefore, the particle can not pass the
barrier and there is no currents. When $T\rightarrow \infty$ so
that the thermal noise is very large, the ratchet effect disappear
and $J\rightarrow 0$, also. There is an optimized value of $T$ at
which the current $J$ takes its maximum value. There is no current
reversal at $\delta=0.1, \Delta=0.9$; $\delta=0.1, \Delta=0.1$;
$\delta=-0.1, \Delta=-0.1$; and $\delta=-0.1, \Delta=-0.9$. In
fact, the temperature cannot lead to the current reversal if
$\Delta\delta>0$ (Fig. 2-Fig. 4).

\begin{figure}[htbp]
  \begin{center}
  \includegraphics[width=8cm,height=8cm]{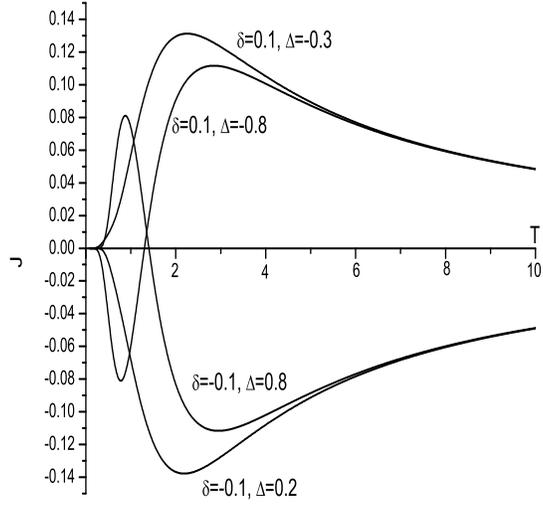} \begin{center}(a)\end{center}
  \includegraphics[width=8cm,height=8cm]{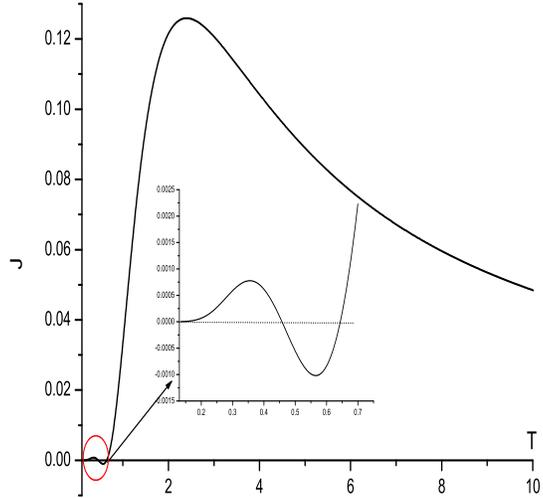} \begin{center}(b)\end{center}
  \caption{\baselineskip 0.4 in (a)Current $J$ versus temperature $T$ for different values
  of the asymmetric parameters $\Delta$ at $U_{0}=5$, $F_{0}=3.0$, $L=1.0$.
  (b)Current $J$ versus temperature $T$ at $U_{0}=5$, $F_{0}=3.0$, $L=1.0$, $\delta=0.1$ and $\Delta=-0.4$.}\label{1}
\end{center}
\end{figure}
\indent In Fig. 5a, we plot the current $J$ as a function of
temperature $T$ for different combinations of $\Delta$ and
$\delta$. When temperature difference ($\delta=0.1$) is positive,
the current may reverse its direction as increasing temperature
for negative $\Delta$ ($\Delta=-0.8$). It is observed that the
current reversal may occur for negative $\delta$ and positive
$\Delta$ ($\delta=-0.1$, $\Delta=0.8$). We can also have the
current reversal twice at $\delta=0.1$, $\Delta=-0.4$ (Fig. 5b).
Therefore, there may exist current reversal for $\Delta\delta<0$.
However, $\Delta\delta<0$ is not sufficient condition for current
reversal. For example, the current is always positive for
$\delta=0.1$, $\Delta=-0.3$ and negative for $\delta=-0.1$,
$\Delta=0.2$ (Fig. 5a).

\begin{figure}[htbp]
  \begin{center}\includegraphics[width=10cm,height=8cm]{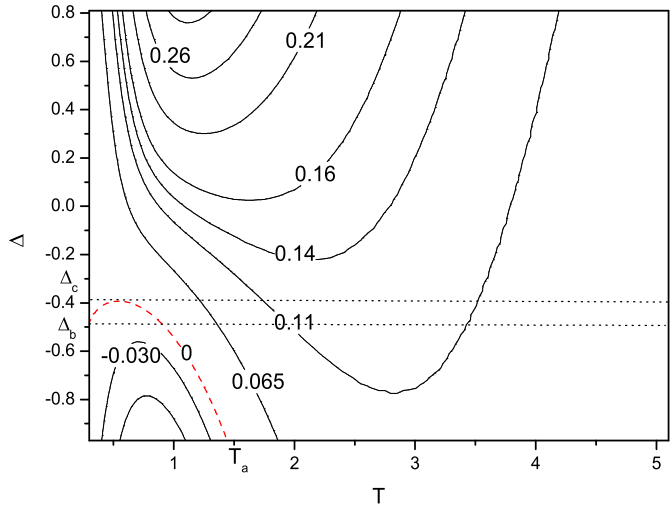}
  \caption{\baselineskip 0.4 in Current contours on  $\Delta$-$T$ plane at $U_{0}=5$, $F_{0}=3.0$, $L=1.0$ and $\delta=0.1$:  $\Delta_{c}$ (-0.3987) is
  maximum $\Delta$ for the curve $J=0$; $\Delta_{b}$ (-0.4913) is asymmetric parameter of
  potential at the cross point of $J=0$ and $T=0$; $T_{a}$ is
  temperature at the cross point of $J=0$ and $\Delta=-1.0$. }\label{1}
\end{center}
\end{figure}

\indent In order to illustrate the current reversal in detail, the
current contours are shown in Figs. 6 and 7, respectively. When
$T>T_{a}$ or $\Delta>\Delta_{c}$,  the current is always positive
 there is no current reversal (Fig. 6, also see the case $\delta=0.1$,
$\Delta=-0.3$ in Fig. 5a). The current reversal may, however,
occur by varying $T$ or $\Delta$ when $T<T_{a}$ or
$\Delta<\Delta_{c}$ (Fig. 6, also see the case $\delta=0.1$,
$\Delta=-0.8$ in Fig. 5a).  In particular, the current may reverse
its direction twice if $\Delta_{b}<\Delta<\Delta_{c}$ (Fig6, also
see the case $\delta=0.1$, $\Delta=-0.4$ in Fig .5b).

\begin{figure}[htbp]
  \begin{center}\includegraphics[width=10cm,height=8cm]{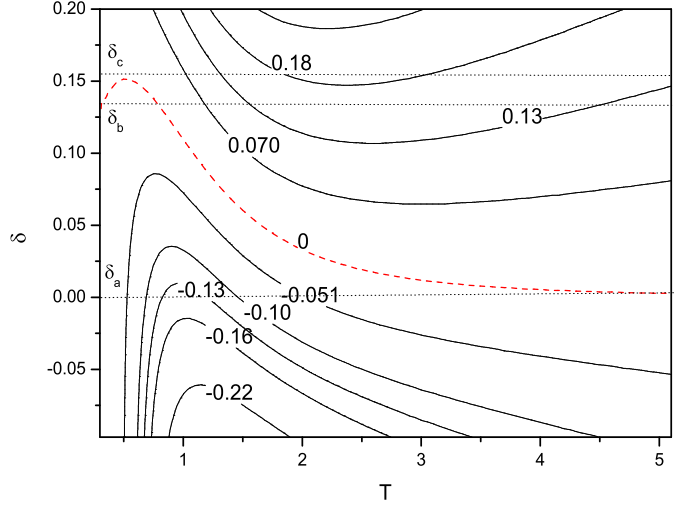}
  \caption{\baselineskip 0.4 in Current contours on $\delta$-$T$ plane at $U_{0}=5$, $F=3.0$, $L=1.0$ and $\Delta=-0.6$: $\delta_{a}=0$; $\delta_{b}$ (0.138) is
  temperature difference at the cross point of $J=0$ and $T=0$; $\delta_{c}$ (0.151) is maximum temperature difference for $J=0$.}\label{1}
\end{center}
\end{figure}

\indent The current is always negative and there is no current
reversal for $\delta\leq 0$ and $\Delta=-0.6$ (Fig. 7). When
$\delta_{a}<\delta<\delta_{c}$, the current may reverse its
direction as increasing temperature. The current may changes its
direction twice, in particular, when
$\delta_{b}<\delta<\delta_{c}$.

\indent Therefore, we can not have the current reversal when
$\delta\Delta\geq0$. When $\delta\Delta< 0$, the current may
reverse its direction. However, $\delta\Delta< 0$ is not the
sufficient condition for current reversal.

\section{4. Concluding Remarks}
\indent The transport of a Brownian particle moving in a periodic
potential is studied in the presence of an unbiased fluctuation
and two heat reservoirs. In a quasi-steady state limit, we obtain
the current analytically. It is found that the asymmetric
parameter $\Delta$ of the potential and the temperature difference
$\delta$ are the two pivotal factors for obtaining a net current.
For the two positive or the two negative driving factors such that
$\Delta\delta>0$, current cannot reverse its direction. The
current reversal cannot occur either if there is only one driving
factor such that $\Delta\delta=0$. For the two opposite driving
factors so that $\Delta\delta<0$, the current reversal may occur.
The current reversal can also occur twice at certain conditions.
The condition $\Delta\delta<0$ is a necessary but not a sufficient
condition for the current reversal.


\begin{thebibliography}{}

\baselineskip 0.4in
\bibitem{1}J. L. Mateos, Phys. Rev. Lett. 84, 258 (2000).
\bibitem{2}J. Luczka, R. Bartussek and P. Hanggi, Europhysics
Letters 31 (8), 431-436 (1995).
\bibitem{3}L. P. Faucheux et al., Phys. Rev. Lett. 74 (1995) 1504.
\bibitem{4}B. Q. Ai, X. J. Wang, G. T. Liu and L. G. Liu, Phys. Rev. E,
68, 061105 (2003); B. Q. Ai, X. J. Wang, G. T. Liu and L. G. Liu,
Phys. Rev. E, 67, 022903 (2003); B. Q. Ai, G. T. Liu, H. Z. Xie,
L.G. Liu, Chaos 14(4),957 (2004).
\bibitem{5}J. Maddox, Nature (London) 368 287(1994); R. D. Astumian and I. Derenyi, Eur. Biophys. J. 27, 474 (1998);
N. Thomas and R. A. Thornhill, J.Phys. D 31, 253 (1998); C.R.
Doering, B. Ermentrout, G. Oster, Biophys. J. 69, 2256(1995);
\bibitem{6}J. Krim, D.H. Solina, R. Chiarello, Phys. Rev. Lett. 66
(1991) 181; J.B. Sokolo2, J. Krim, A. Widom, Phys. Rev. B 48
(1993) 9134; L. Daikhin, M. Urbakh, Phys. Rev. E 49 (1994) 1424;
C. Daly, J. Krim, Phys. Rev. Lett. 76 (1996) 803; M.R. Sorensen,
K.W. Jacobsen, P. Stoltze, Phys. Rev. B 53 (1996) 2101.
\bibitem{7}I. Derenyi, C. Lee, and A.-L. Barabasi, Phys. Rev. Lett. 80 (1998)
1473.
\bibitem{8}I. Zapata, R. Bartussek, E. Sols, P. HFanggi, Phys. Rev. Lett.
77, 2292 (1996).
\bibitem{9}L.P. Faucheux, L.S. Bourdieu, P.D. Kaplan, A.J. Libchaber, Phys. Rev. Lett.
74,1504 (1995); C. Mennerat-Robilliard, D. Lucas, S. Guibal, J.
Tabosa, C. Jurczak, J.-Y. Courtois, G. Grynberg, Phys. Rev. Lett.
82, 851(1999).
\bibitem{10}L. Gorre-Talini, J.P. Spatz, P. Silberzan, Chaos 8, 650 (1998);
I. Derenyi, R.D. Astumian, Phys. Rev. E 58, 7781 (1998); D. Ertas,
Phys. Rev. Lett. 80, 1548 (1998); T.A.J. Duke, R.H. Austin, Phys.
Rev. Lett. 80, 1552 (1998).
\bibitem{11}M. O. Magnasco, Phys. Rev. Lett. 71,  1477 (1993).
\bibitem{12}P. Hanggi and R. Bartussek, Nonlinear physics of
complex system - Current status and Future Trends, 476, Spring,
Berlin, (1996), 294.
\bibitem{13}P. Reimann, R. Bartussek, R. Haussler and P. Hanggi,
Phys. Lett. A  215,  26 (1994).
\bibitem{14}C. R. Doering, W. Horsthemke and J. Riordan, Phys. Rev.
Lett. 72,  2984 (1994).
\bibitem{15}C. kettner, Phys. Rev. E 61, 312 (2000).
\bibitem{16}R. Tammelo, R. Mankin and D. Martila, Phys. Rev. E 66,
051101 (2002).
\bibitem{17}D. Dan, M. C. Mahato and A. M. Jayannavar, Phys. Rev.
E 63, 056307 (2001).
\bibitem{18}B. Q. Ai, X. J. Wang, G. T. Liu, H. Z. Xie, D. H.  Wen,  W. Chen, and L. G. Liu., Eur. Phys. J.
B 37, 523-526 (2004).
\bibitem{19}M. M. Millonas and M. I. Dykman, Phys. Lett. A 185, 65
(1994).
\bibitem{20}J.-F. Chauwin, A. Ajdari, and J. Prost, Europhys. Lett.
32, 373 (1995).
\bibitem{21}M. Bier and R. D. Astumian, Phys. Rev. Lett. 76,
4277 (1996).
\bibitem{22}C. R. Doering, W. Horsthemke, and J. Riordan, Phys. Rev.
Lett. 72, 2984 (1994).
\bibitem{23}R. Bartussek, P. Reimann, and P. Ha¡§nggi, Phys. Rev. Lett. 76,
1166 (1996).
\bibitem{24}R. Mankin, A. Ainsaar, A. Haljas and E. Reiter, Phys.
Rev. E 63, 041110 (2001).
\bibitem{25}J. Kula, T. Czernik and J. Luczka, Phys. Rev. Lett.
80, 1377 (1998).
\bibitem{26}M. Kostur and J. Luczka, Phys. Rev. E 63, 021101
(2001).
\bibitem{27}I. Derenyi and A. Ajdari, Phys. Rev. E 54, R5 (1996).
\bibitem{28}F. Marchesoni, Phys. Lett. A 237, 126 (1998);
H. A. Larrondoa, Fereydoon Family, C. M. Arizmendia, Physica A
303, 67-68 (2002).
\bibitem{29}Oster, G., H. Wang. How protein motors convert chemical energy into mechanical work.
In Molecular Motors, M. Schliwa, ed. pp. 207-228. ISBN 3-527-30594-7. Wiley-VCH (2002).
\bibitem{30}H. Risken, The Fokker-Planck Equation, Springer-Verlag, Berlin, 1984.
\bibitem{31}M. Asfaw and M. Bekele, Eur. Phys. J. B 38, 457 (2004).
\bibitem{a}U. Henningsen, M. Schliwa, Nature 389, 93 (1997)
\end{thebibliography}
\end{document}